\pdfoutput=1

\documentclass[3p,times,twocolumn]{elsarticle}

\usepackage{ecrc}

\volume{00}

\firstpage{1}

\journalname{Nuclear Physics B Proceedings Supplement}

\runauth{}

\jid{nuphbp}

\jnltitlelogo{Nuclear Physics B Proceedings Supplement}

\usepackage{amssymb}
\usepackage[figuresright]{rotating}

\begin{document}

\begin{frontmatter}

\dochead{}

\title{Cosmic rays: interstellar gamma-ray and radio emission}

\author[label1]{Elena Orlando}
\author[label2]{Andrew Strong}
\address[label1]{W.W. Hansen Experimental Physics Laboratory, Kavli Institute for Particle Astrophysics and Cosmology, Stanford University, Stanford, CA, 94305, USA}
\address[label2]{Max-Planck-Institut fuer extraterrestrische Physik, Postfach 1312, 85741, Garching, Germany}

\begin{abstract}
We discuss the relation between cosmic-ray induced interstellar gamma-ray and radio emission from our Galaxy, 
and emphasize the importance of their parallel study.
We give  an overview of  results from  Fermi-LAT on the
interstellar gamma-ray emission, and then  focus on complementary studies
of the radio emission from the Galaxy. 
\end{abstract}

\begin{keyword}

Interstellar gamma rays; interstellar radio emission - Cosmic rays 
\end{keyword}

\end{frontmatter}

\section{Introduction}
Cosmic rays (CRs) propagate in the Galaxy and interact with the interstellar medium (ISM). From their interactions with the gas and the interstellar radiation field (ISRF), gamma-ray diffuse emission is produced. The mechanisms of this diffuse emission are: neutral pion decay from hadronic interactions of CR protons and heavier nuclei
with gas, inverse-Compton (IC) scattering from CR electron and positron interactions with the ISRF and the cosmic microwaves background (CMB), and bremsstrahlung from CR electron interactions with gas (see eg. \cite{diffuse1}). 
The same CR electrons and positrons producing gamma rays also generate  diffuse radio emission, via synchrotron radiation in  Galactic magnetic fields (B-field). 
Interstellar synchrotron emission extends from a few MHz to tens of GHz and is one of the major components of radio and microwave band radiation. 
This emission depends  on the CR electron spectrum and distribution in the Galaxy, and on the B-field (see eg. \cite{strong2011}). 

Observations of gamma-ray and radio emission from our Galaxy are an important tool for studying CR, the  ISRF, interstellar gas and the B-field. 
Using the combination of these energy bands and direct measurements of local CRs, we can gain information on CR propagation models, CR source distribution, and CR spectra in  interstellar space.
The importance of parallel studies of radio and gamma-ray emission was recognized already in the 80's, eg. in  \cite{haslam81}, \cite{higdon79}, \cite{strong78} and \cite{strong2000} and more recently in \cite{orlando2012a}, \cite{strong2011}. 
This approach has advantages  over  studying these data separately, since the CR electrons are traced by both gamma-ray and radio emissions and so degeneracy is removed. 
CR electrons are measured  in direct observations, but these are affected by solar modulation at lower energies.

We present here an overview of the latest results with
Fermi-LAT on the gamma-ray diffuse emission induced by CR nuclei and electrons.
 Then we
focus on the complementary studies of the synchrotron emission in the light of the latest
gamma-ray results.
These studies made use of the  numerical model of CR propagation and interactions in the Galaxy, GALPROP \citep{moskalenko98, strong98}, described in Section 3. 

\label{Introduction}


\section{Cosmic rays throughout the Galaxy}

CRs are believed to originate in supernovae remnants (SNRs). During propagation in the Galaxy, the CR spectra are steepened  by  energy losses and energy-dependent diffusion, but  can also be affected  at low energies by diffusive reacceleration. 
The spectrum  originating in the sources is called the injected spectrum,  the propagated spectrum pertains to interstellar space, while that measured in the Solar System is the solar-modulated spectrum. 
Inverse-Compton scattering, synchrotron radiation and ionization are responsible for energy losses of electrons and positrons. 
Direct measurements of CRs by balloon experiments and satellites constrain models of propagation and local CR spectra. 
However solar modulation complicates the interpretation of the observations at energies below about 10 GeV/n. 
An important point is that gamma-ray and synchrotron emission probe the interstellar CR spectra free from  the effects of  modulation. 


\section{Modelling interstellar emission with GALPROP}

GALPROP  is a software package for modelling the propagation of CR in the Galaxy and their interactions in the ISM. 
Descriptions of the GALPROP software can be found in \cite{moska2000}, \cite{strong2004}, \cite{strong2007} and \cite{vladimirov} and references therein. 
This project started in the late 1990's \citep{moskalenko98, strong98}. 
Since then, it has been continuously improved and updated.
It enables simultaneous
predictions of observations of CRs, gamma rays and recently synchrotron radiation \citep{orlando2009,  orlando2012b, strong2011}.
 GALPROP models were used to analyse the diffuse gamma rays detected by CGRO EGRET and COMPTEL, INTEGRAL \citep{porter2008}, and more recently radio surveys and WMAP data \citep{orlando2012b, strong2011}. 
It is used in the Fermi-LAT collaboration for interpreting the physics of the Galactic emission \cite{diffuse2}.
For details we refer the reader
to the relevant papers \citep{diffuse2, strong2011} and the dedicated website\footnote{http://galprop.stanford.edu/}. 
GALPROP solves the transport equation for a given CR  source distribution and boundary conditions, for all required CR species. 
It takes into account diffusion, convection, energy losses, and diffusive reacceleration processes. 
Secondary CRs produced by collisions in the ISM and decay of radiative isotopes are included.  
The propagation equation is solved numerically on a user-defined spatial grid in 2D or in 3D, and energy grid. 
The solution proceeds until a steady-state solution is obtained, starting with the heaviest primary nucleus, and then  electrons,
positrons, and antiprotons are computed. 
GALPROP models gamma-ray emission of the three components, pion decay, bremsstrahlung and Inverse Compton for a user-defined CR source distribution and CR spectra. 
Gas maps and ISRF are provided and updated to the most recent observations \citep{diffuse2}. 
More recently, GALPROP calculation  of interstellar synchrotron emission has been improved \citep{strong2011} and it has been extended to include also  synchrotron polarization, and free-free emission  and  absorption \citep{orlando2012b}. 

\begin{figure}
\centering
\includegraphics[width=18pc, angle=0]{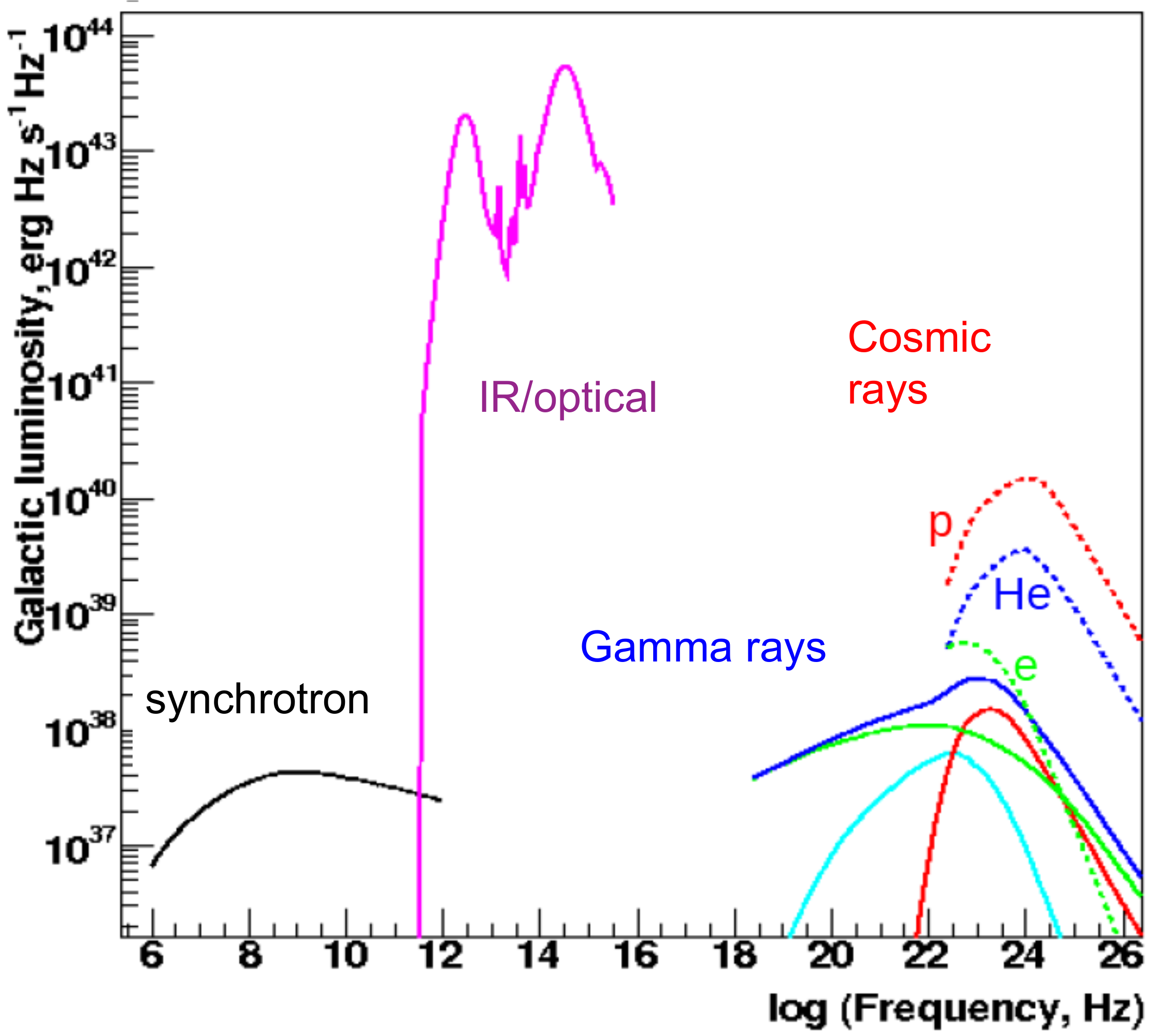}
\caption{Global luminosity spectra of the Galaxy. Line styles: interstellar radiation field, including optical and infrared  (magenta solid), protons (red dotted line), helium (blue dotted line), primary electrons (green dotted line); CR-induced diffuse emissions (solid lines): IC (green),
bremsstrahlung (cyan), $\pi ^0$ decay (red), synchrotron (black), total (blue).
CR particle luminosities are also shown as labelled.}
\label{fig1}
\end{figure}

In \cite{strong2010} we used the GALPROP code to derive the broad-band CR-induced luminosity spectrum
of the Galaxy. Figure \ref{fig1} shows an example of the modelled luminosity of the Galaxy.
The total luminosity in synchrotron and inverse Compton emission are similar in magnitude.
We found the Galaxy to be a fairly good lepton calorimeter, when both inverse Compton and synchrotron losses are accounted for.

\section{Interstellar gamma-ray emission}

Interstellar gamma-ray emission accounts for at least  60\% of the gamma-ray photons detected by Fermi-LAT.  
An early result from Fermi-LAT \cite{diffuse1} was a study of this  gamma-ray emission. 
This study was carried out using data from intermediate Galactic latitudes, where the interstellar emission comes mainly from local CR interactions
with the ISM.
The spectrum was found to be well described by pion decay, IC and bremsstrahlung, plus the contribution from point
sources and an isotropic component, which included the extragalactic background. 
This result was confirmed in \cite{thirdquad} and  \cite{secondquad}, where the local HI gamma-ray emissivity was derived for the 2nd and 3rd
Galactic quadrants.
The latter  studies also showed an increase in the H$_2$-to-CO factor (X$_{CO}$) with
Galactocentric radius and larger intensities in the outer Galaxy with respect to the a priori model based on local CR measurements. 
A flatter distribution of CR sources in the outer Galaxy or a larger amount of gas could explain this effect. 
Models used for these papers, generated with GALPROP, included diffusive reacceleration with a 4 kpc
halo height,  the locally-measured CR proton and Helium spectra, and the electron spectrum as measured by Fermi \cite{fermiel1} and \cite{fermiel2}. 
Models used Alfv\'en velocity in the range 30-39~km~s$^{-1}$ (i.e. see \cite{diffuse2}, \cite{strong2007}, \cite{strong2010}, \cite{Trotta}, \cite{vladimirov}).
CR sources were assumed to follow supernova remnants as traced by pulsars.
A preliminary study including the Galactic plane was presented in \cite{strongICATPP}, where the spectrum of the inner Galaxy, longitude
and latitude profiles were shown. The model reproduced the Fermi data over the sky within 20$\%$; the inner Galaxy revealed an
excess at GeV energies that can plausibly be explained by unresolved sources. It was shown that latitude profiles improved with a
larger halo height of about 10 kpc, consistent with the previous works mentioned above. 
More recently \citep{diffuse2} investigated models of the gamma-ray diffuse emission, taking into account uncertainties associated
with the astrophysical parameters. 
The distribution of cosmic-ray sources, the size of the cosmic-ray halo, and the interstellar
gas distribution were varied within observational limits. 
They produced a grid of  GALPROP models and compared them with Fermi-LAT data for the first 21 months of observations. 
They found a general  agreement between  models and Fermi-LAT data, showing that the physics of the diffuse
emission is  broadly understood.
The gamma-ray fits improved using dust as a tracer of gas,  confirming  the presence of molecular hydrogen that is not traced
by CO: the so-called dark gas as found by \cite{grenier} and recently confirmed by \cite{planck}.
No single model gave a best fit over all sky regions.
In general, the models were consistent with the data at high and intermediate latitudes.
However, all models underpredicted the data in the inner Galaxy for energies above a few GeV. 
In the outer Galaxy, they found that the data
prefer models with a flatter distribution of cosmic-ray sources or greater gas density. 
In general, all models were within 15\% of data, except for some regions
such as Loop~I,  and the structures coincident with the bubbles found by \cite{su}, and
the above-mentioned excess in the outer Galaxy.


\section{Interstellar radio emission}

Interstellar synchrotron radiation is the major contributor to the Galactic radio emission from tens of MHz to a few GHz. At higher frequencies and in the microwave bands, free-free and dust emission tend to dominate. Synchrotron emission is also a foreground component for CMB studies.  
Recent studies of the Galactic synchrotron emission using CR electron measurements were published in \cite{orlando2011, strong2011}. 
Below a few GeV the local interstellar electron spectrum cannot be directly measured, because CR electrons with energy lower than a few GeV are affected by solar modulation.  
The synchrotron spectral index depends on the CR electron and positron spectral index, while synchrotron intensity depends on the B-field intensity and electron flux.
In particular CR electrons from 500 MeV to 20 GeV produce synchrotron emission from tens of MHz to hundreds of GHz for a B-field of few $\mu$G, and hence this can be used in conjunction with direct measurements to construct the full interstellar electron spectrum from GeV to TeV. 

\begin{figure}
\centering
\includegraphics[width=18pc, angle=0]{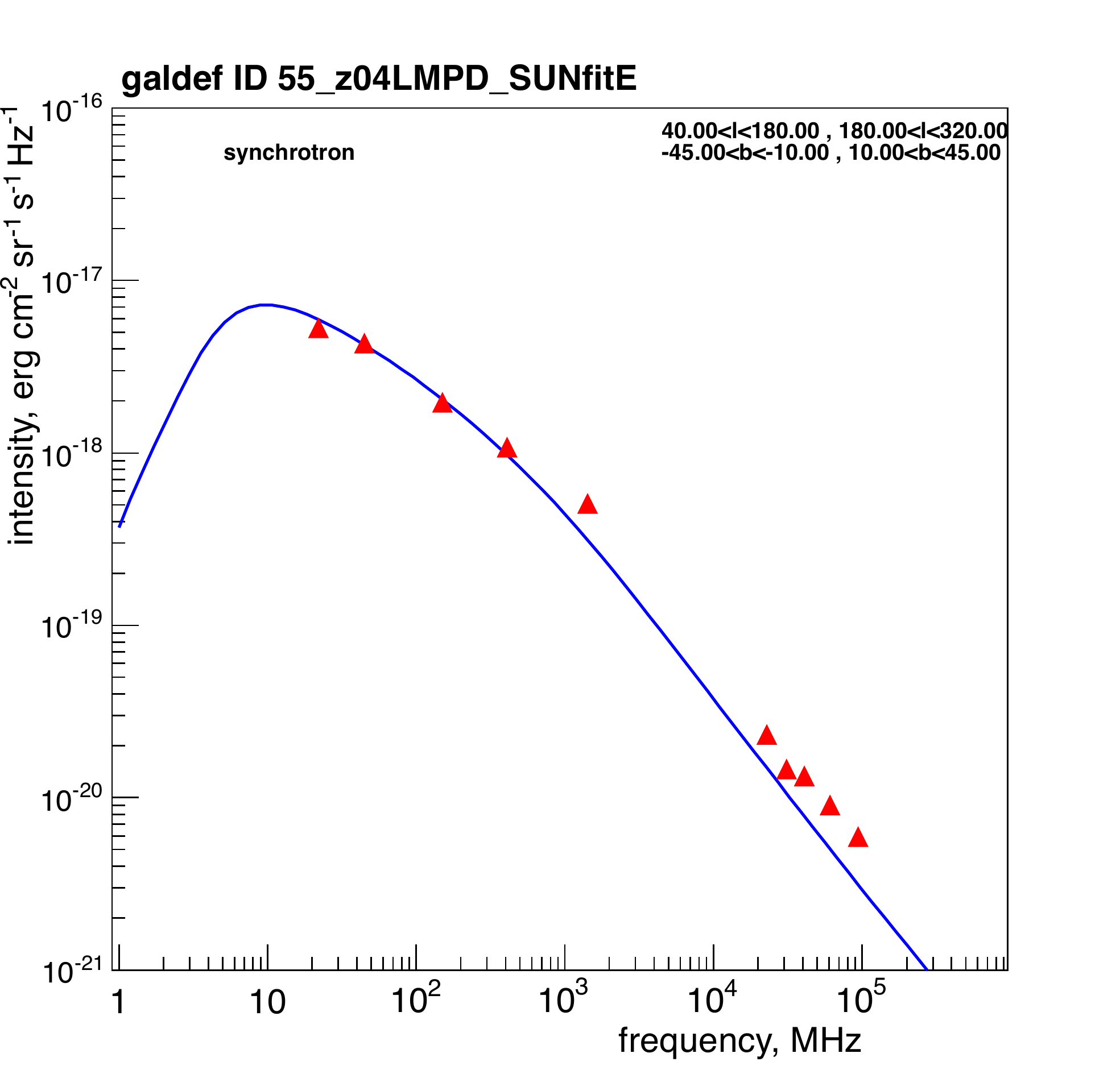}
\includegraphics[width=18pc, angle=0]{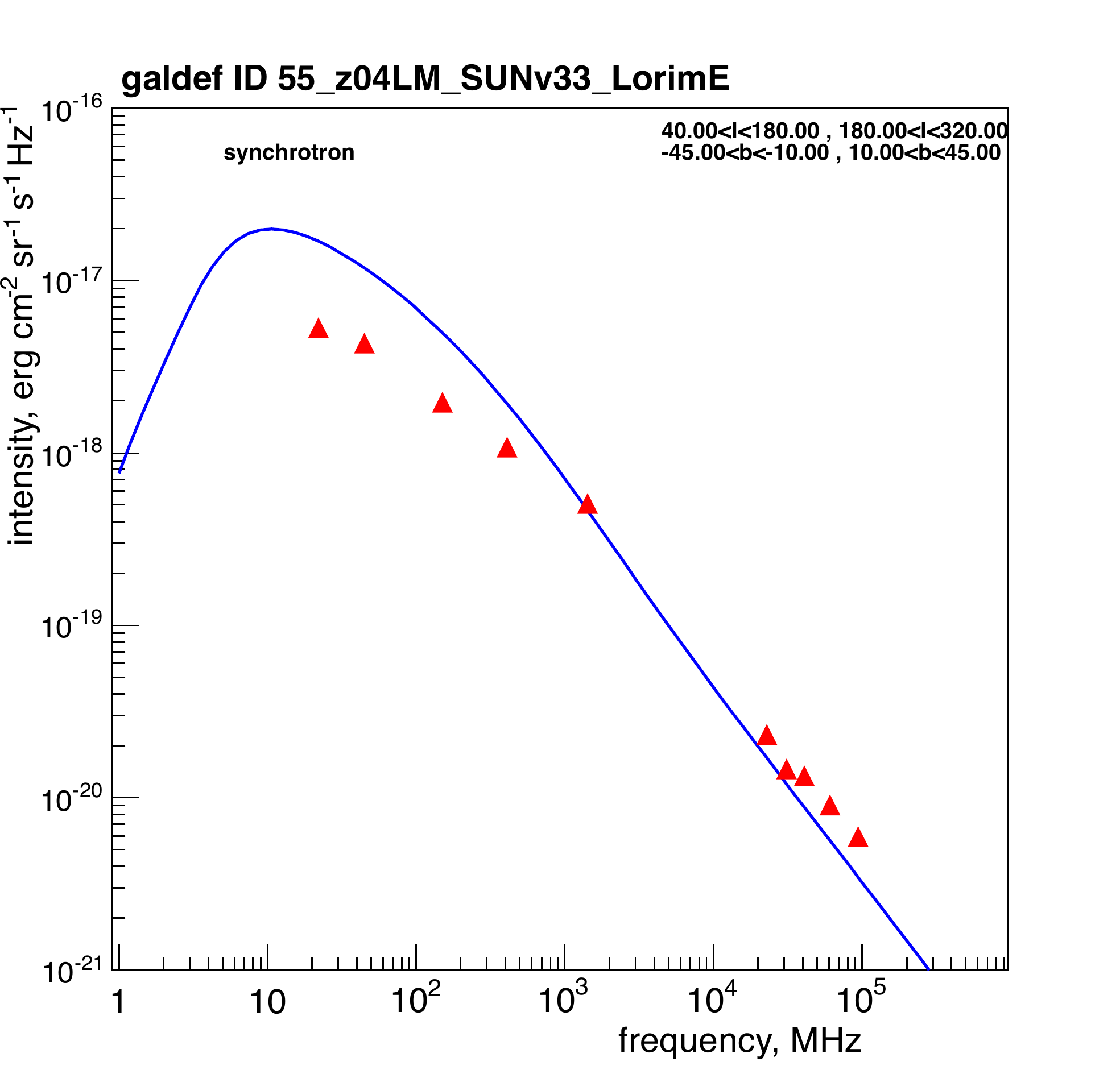}
\includegraphics[width=18pc, angle=0]{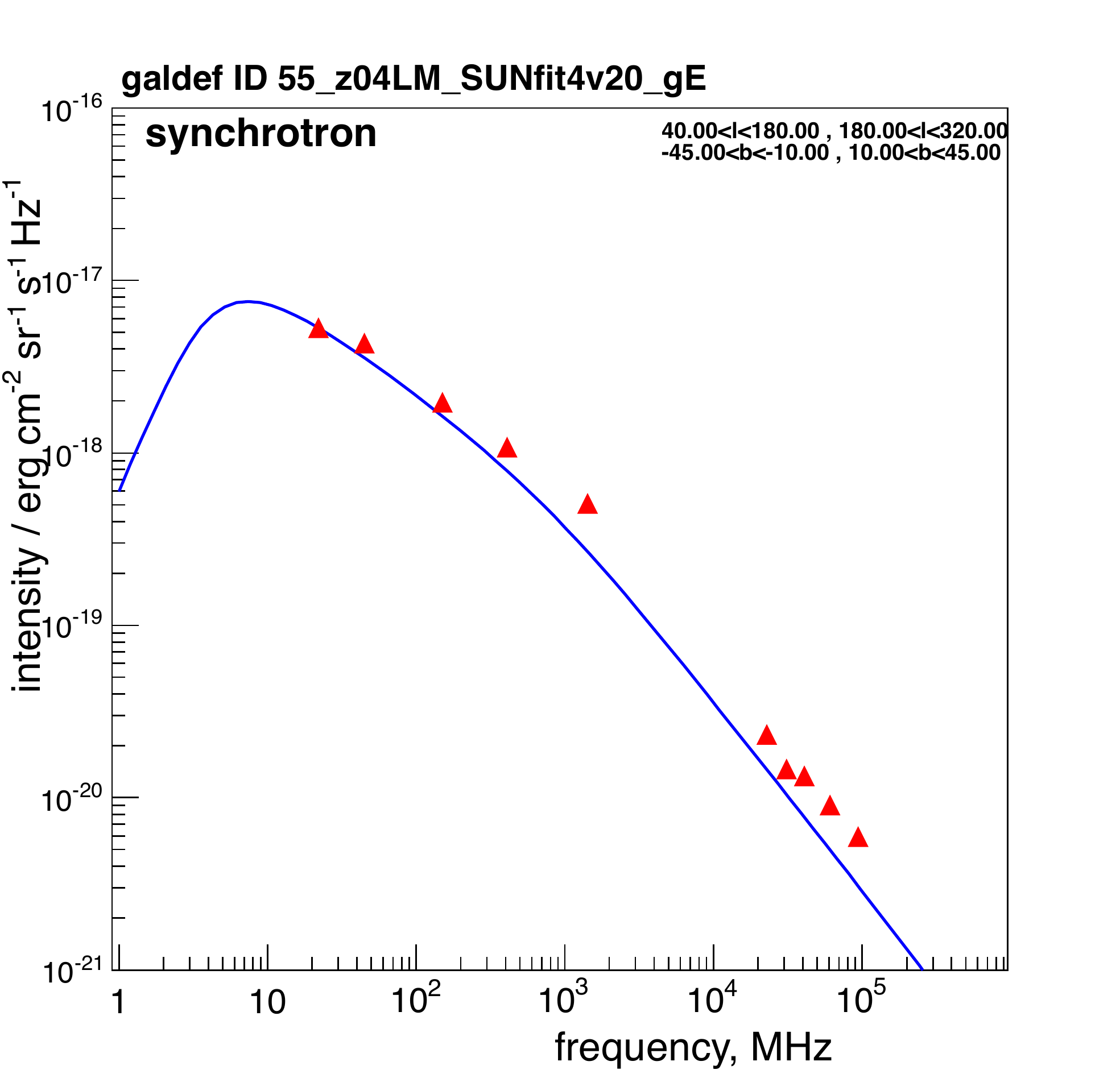}
\caption{Synchrotron spectra (blue lines) with primary low-energy electron injection index 1.6 for pure diffusion model (upper), for diffusive acceleration model as in \cite{diffuse1} (center) and for diffusive reacceleration model with Alfv\'en velocity 20~km~s$^{-1}$ (lower) at high latitudes. Data (red triangles) are from radio surveys  and WMAP.}
\label{fig2}
\end{figure}

\begin{figure}[!h]
\centering
\includegraphics[width=0.45\textwidth, angle=0] {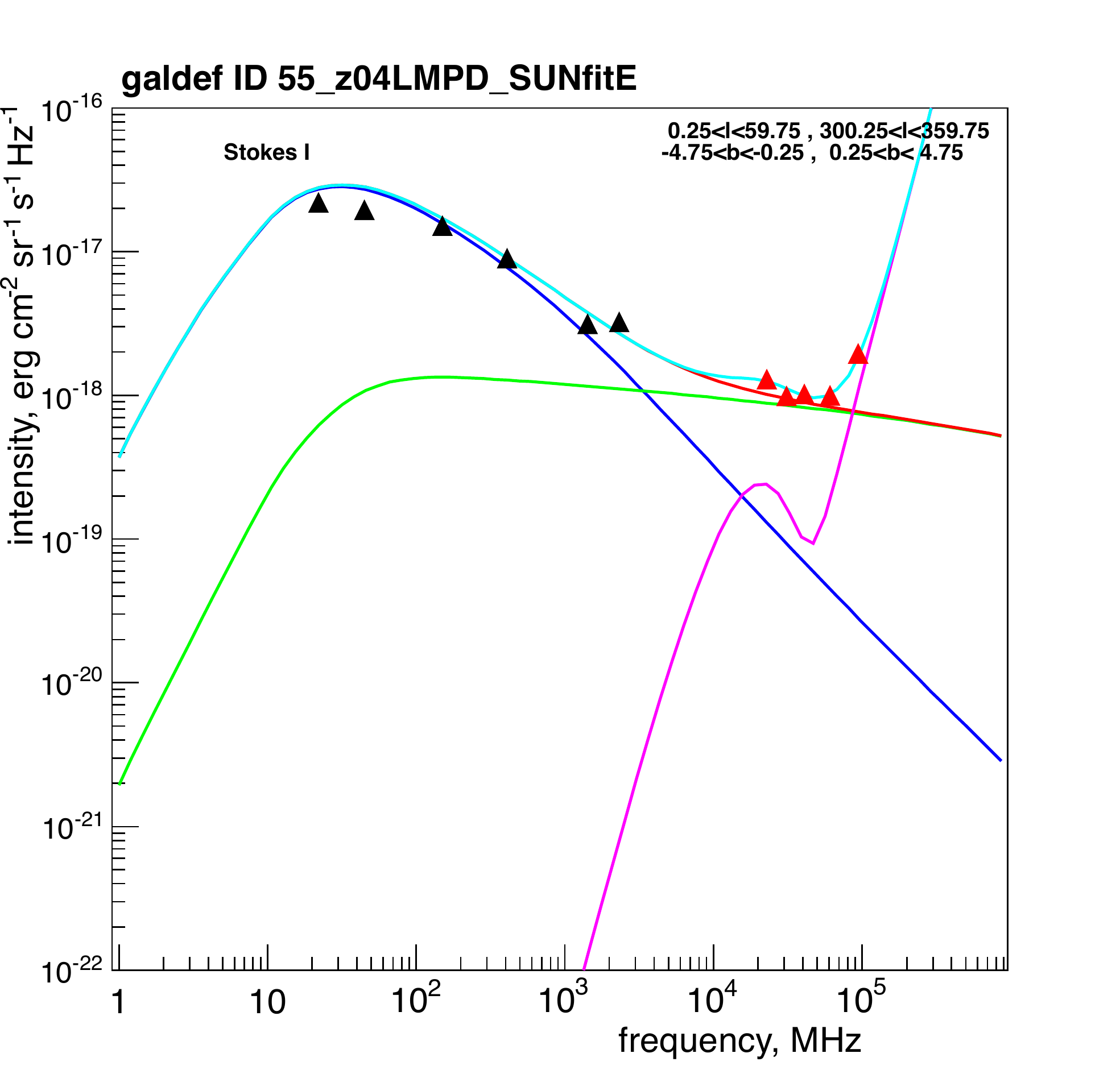}
\includegraphics[width=0.45\textwidth, angle=0] {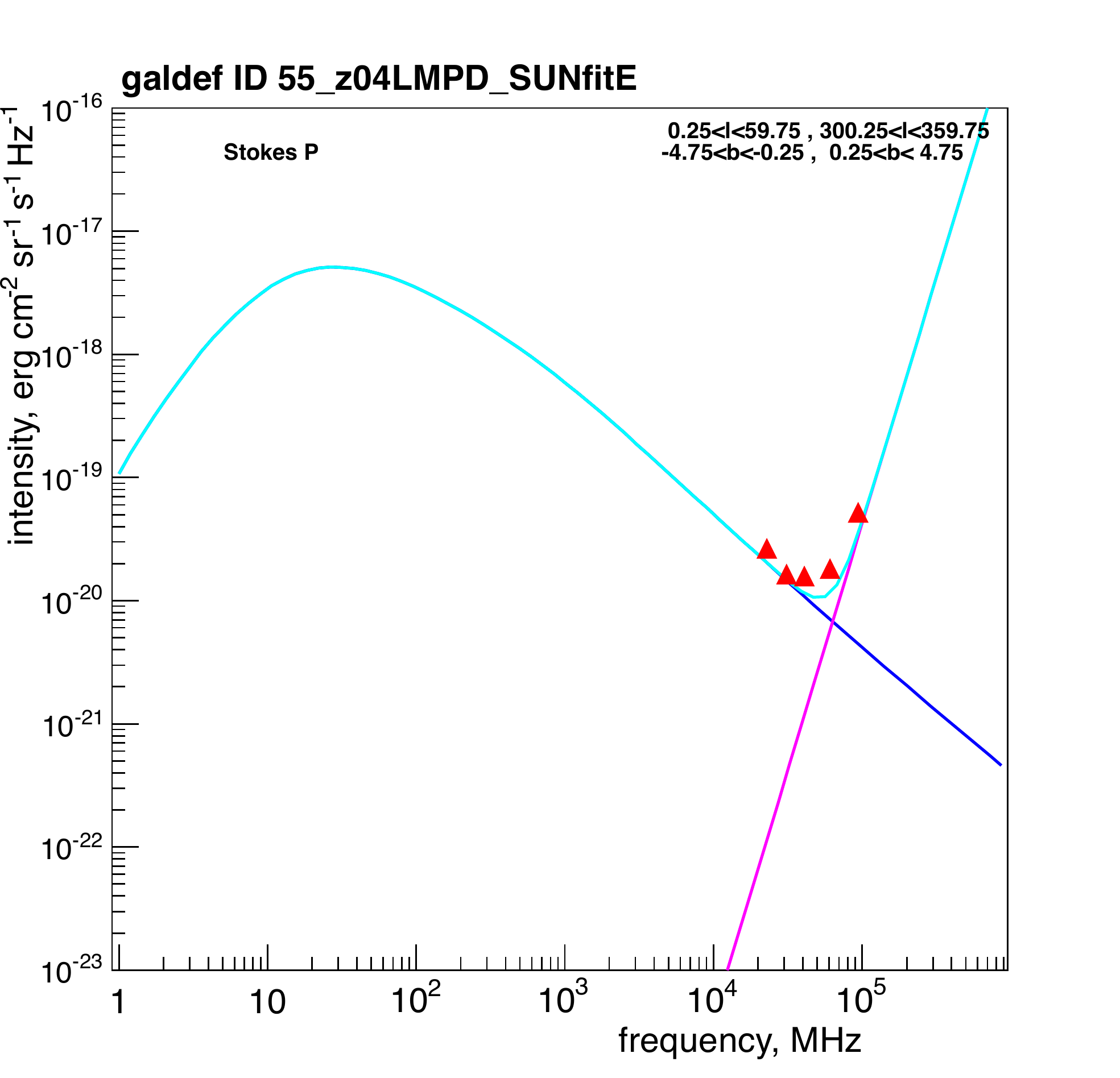}
\caption{  Radio spectra of total intensity I (upper) and polarized intensity P  (lower) for the inner Galaxy. B-field intensities are scaled with respect to the original models to agree with data. The plots show various model components: synchrotron (blue), thermal dust + spinning dust (magenta upper plot),
thermal dust (magenta lower plot), free-free (green), free-free+synchrotron (red line) and total (cyan). Data are from radio surveys (black triangles) and WMAP (red triangles). }
\label{fig3}
\end{figure}

\begin{figure}[!h]
\centering
\includegraphics[width=0.45\textwidth, angle=0] {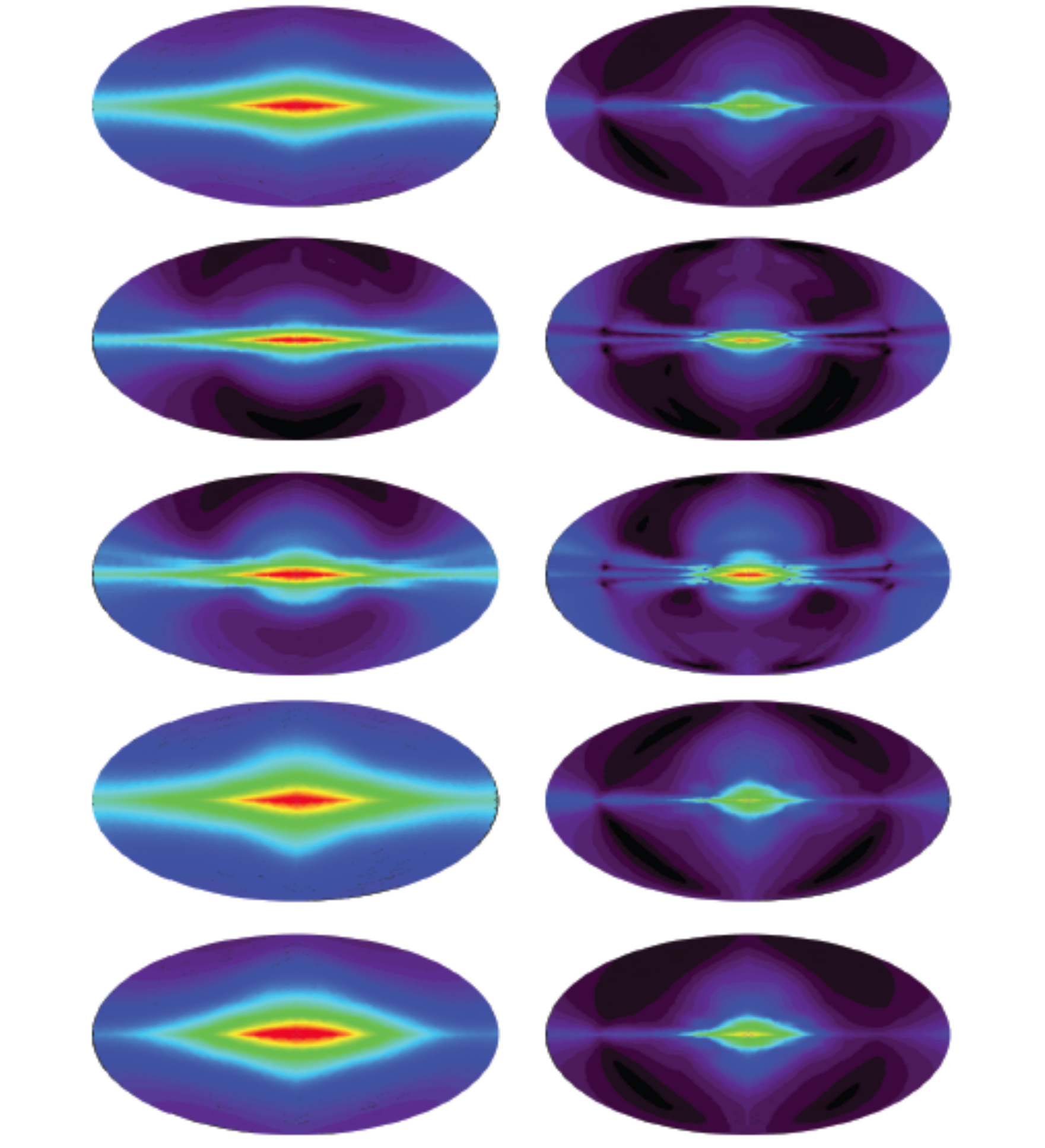}
\includegraphics[width=0.45\textwidth, angle=0] {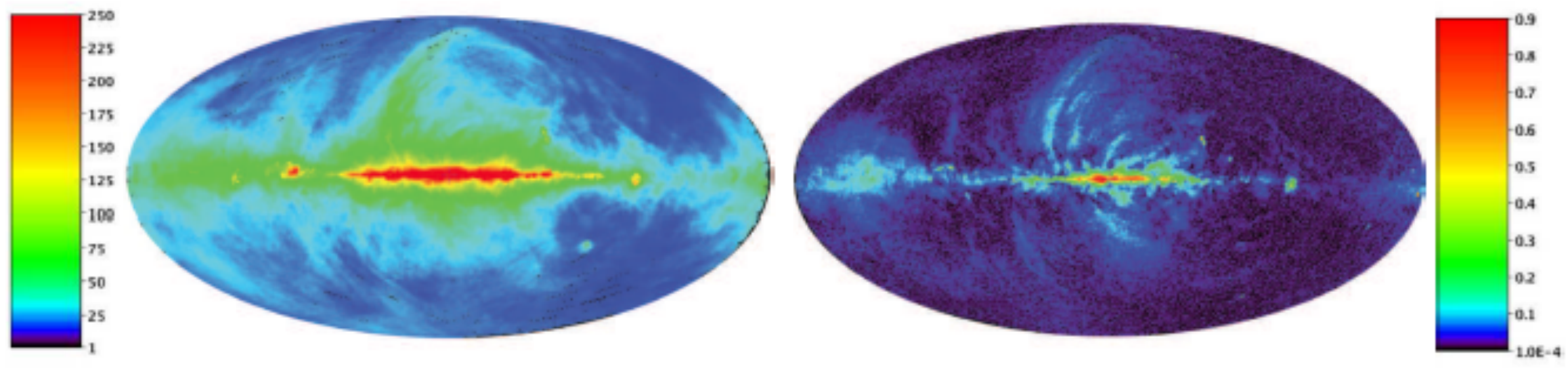}
\caption{Brightness temperature skymaps at 408~MHz  (left) and polarized  23~GHz (right). Top to bottom are the following models: - ASS+RING B-field model based on \cite{sun2008}, for CR source distribution from \cite{strong2010}, for 4 kpc halo size; - ASS B-field model based on \cite{pshirkov}, for CR source distribution from \cite{strong2010}, for 4 kpc halo size; - BSS B-field model based on \cite{pshirkov} , for CR source distribution from \cite{strong2010}, for 4 kpc halo size; ASS+RING B-field model based on \cite{sun2008}, CR source distribution from \cite{strong2010}, for 10 kpc halo size; -  ASS+RING B-field model based on \cite{sun2008}, CR source distribution from \cite{lorimer}, for 4 kpc halo size. The intensities of the local magnetic field are increased of a factor of 2 with respect the original models in order to fit the radio data (see text for more details). For comparison for all the models shown the intensity of the local regular component is 4 $\mu$G and for the random is 6 $\mu$G. The skymaps at the bottom are the data from \cite{haslam81} and WMAP \cite{gold}. All skymaps (for each frequency) have the same scale, with  Galactic longitude l = 0 at the center. Units are K (left) and mk (right).}
\label{fig4}
\end{figure}

Using a collection of radio surveys and WMAP  data, out of the Galactic plane, \cite{strong2011} found that the propagated local interstellar electron spectrum turns over below a few GeV, with an ambient interstellar index around 2, independent of modulation. 
We tested propagation models, generated with GALPROP, in order to constrain the injected CR electron spectrum, before propagation. We found the need for an injection electron spectral index between 1.3 and 1.6 below 4 GeV (see Fig~\ref{fig2}).
One  result was the recognition of the importance of including secondary positrons and electrons in synchrotron models.  
While plain diffusion models fitted the data well, standard reacceleration models often used to explain CR secondary-to-primary ratios were not consistent with the observed synchrotron spectrum, since the total intensity from primary and secondary leptons exceeded the measured synchrotron emission at low frequencies, as shown in  Fig~\ref{fig2}. 
We confirm here that plain diffusion models or diffusive reacceleration models with small Alfv\'en velocity are preferred  to existing reacceleration models, as found  previously in \cite{strong2011}.
Spectra of the total synchrotron emission for high latitudes regions and for the case of lower Alfv\'en velocity than usually assumed are shown in  Fig~\ref{fig2}.
Reacceleration models with Alfv\'en velocity around  20~km~s$^{-1}$ also fit the synchrotron data. 
A detailed description of our recent improvements in the radio modelling is given in \cite{orlando2012b}. 
 An example of our modelling of the radio emission is shown on Figure \ref{fig3}. Here spectra of total (I) and polarized (P) emission  from radio surveys and WMAP data are compared with our model in the inner Galaxy, showing reasonable agreement. 
Polarized synchrotron emission allows the regular component of the B-field to be studied, complementing the total synchrotron which probes the sum of random and regular fields. 
Models of the regular B-field from the literature can be investigated, using CR propagation models based on recent CR measurements. 
We studied the sensitivity of the synchrotron model to various  formulations of the regular B-field in the literature (see caption to Fig~\ref{fig4} for details of the models), finding that  while all models resemble the data to first order,
no single model of the ones used fitted best all the data for the whole sky. 
Skymaps at 408~MHz and polarized emission at 23~GHz are shown in Fig~\ref{fig4} for various models of B-field, halo size, and CR source distribution.
Note that local features like Loop~I are not included in the model, and these account for much of the additional structure seen in the data.
The models used reproduce the peak in the direction of the inner Galaxy for both total and polarized radio components.  
For the magnetic field models used, 3--4~$\mu$G and 5--6~$\mu$G respectively for the regular and random component of the local B-field described the data best.
This regular field is a factor 1.5--2 larger than in the original models.
We find that increasing the height of the halo from 4~kpc to 10~kpc gives better agreement with radio data for high latitudes. However it somewhat overestimates the emission at intermediate latitudes. Hence no conclusions can be drawn at this point, and the halo height should be investigated in more detail  in the future.  
The effect of varying the CR source distribution has also been investigated, indicating that a flat distribution in the outer Galaxy reproduces better the data.

\begin{figure}
\includegraphics[width=18pc, angle=0]{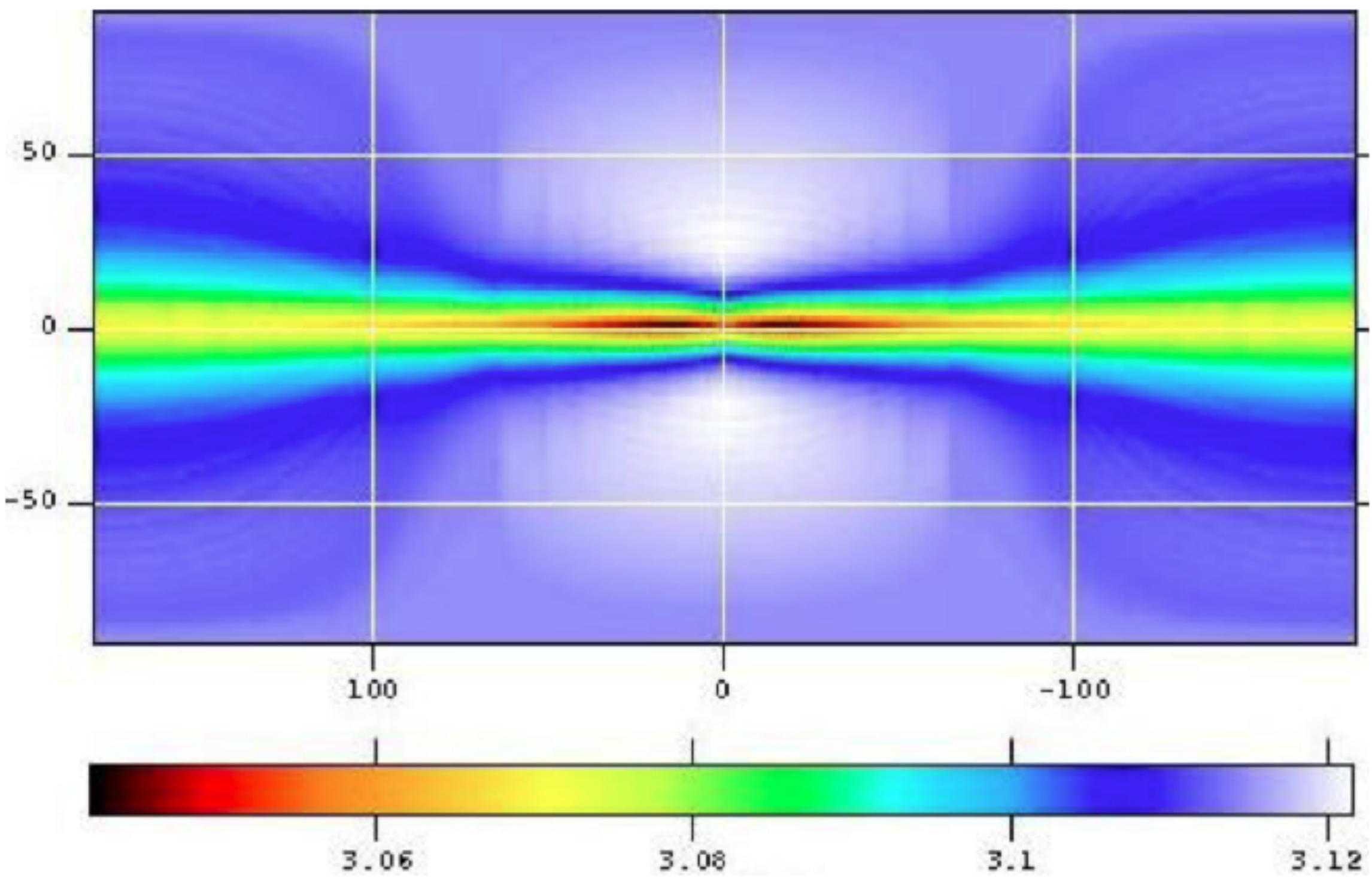}
\caption{ An example of the modelled synchrotron spectral index from 408 MHz to 23 GHz for the full sky. The map is in Galactic coordinates.}
\label{fig5}
\end{figure}

We model also the expected synchrotron spectral index and its variation in the Galaxy.  
Figure \ref{fig5} shows an example of a skymap of the spectral index from 408 MHz to 23 GHz, for a given model of magnetic field. We see that the variations in the Galaxy are very small compared for example with the one obtained by \cite{bennett} with WMAP where the spectral index range is 2.3-3.4.  However the trend is similar, with a smaller spectral index in the plane, and increasing away from the plane. This means that the physical model of electron propagation is supported.


\section{Summary}

We have described some recent  results on   interstellar emission related to Galactic CR. We have also presented the recent improvements on our modelling of the interstellar radio emission.
The aim  of this paper is to highlight  what we can learn from  a parallel study of  gamma rays and radio emission in a multi-wavelength approach. 

Generally  the physics of these interstellar emissions is well understood.  
Models with a CR source distribution rather flat in Galactocentric radius beyond 10 kpc  reproduce better both gamma-ray and radio data in the outer Galaxy. 
A halo size larger than 4 kpc is indicated in both energy bands in order to fit the high latitude regions. 
However, no single model is found to  reproduce the data best over the sky. 
Present diffusive reacceleration models, with parameters often used to reproduce secondary/primary ratios, are not consistent with synchrotron data.
In order to reproduce synchrotron data, plain diffusion models or  models with small diffusive reacceleration are needed. 
Gamma-ray studies alone leave unanswered questions which can be addressed by investigating the radio band, hence putting more constraints on the models. 
For example, gamma-ray data are not very sensitive to the electron spectrum, because gamma emission is dominated by the hadronic component rather than the IC  emission. In this case, gamma rays and radio data give complementary information.

These studies are also of interest for upcoming data from the Planck mission.\\
\\

Acknowledgments: 
This work makes  use of HEALPix\footnote{http://healpix.jpl.nasa.gov/} described in \cite{healpix}.
E.O. acknowledges support 
via NASA Grant No. NNX13AH72G and NNX09AC15G.


\nocite{*}
\bibliographystyle{elsarticle-num}



\end{document}